\magnification \magstep1
\raggedbottom
\openup 2\jot
\voffset6truemm
\def\cstok#1{\leavevmode\thinspace\hbox{\vrule\vtop{\vbox{\hrule\kern1pt
\hbox{\vphantom{\tt/}\thinspace{\tt#1}\thinspace}}
\kern1pt\hrule}\vrule}\thinspace}
\def\II{{\rm1\!\hskip-1pt I}}
\centerline {\bf Euclidean Quantum Gravity in Light of
Spectral Geometry}
\vskip 1cm
\centerline {Giampiero Esposito}
\vskip 1cm
\noindent
{\it INFN, Sezione di Napoli, Complesso Universitario di Monte
S. Angelo, Via Cintia, Edificio N', 80126 Napoli, Italy}
\vskip 0.3cm
\noindent
{\it Dipartimento di Scienze Fisiche, Universit\`a Federico II
di Napoli, Complesso Universitario di Monte S. Angelo, Via
Cintia, Edificio N', 80126 Napoli, Italy}
\vskip 1cm
\noindent
{\bf Abstract}. A proper understanding of boundary-value problems
is essential in the attempt of developing a quantum theory of
gravity and of the birth of the universe. The present paper 
reviews these topics in light of recent developments in spectral
geometry, i.e. heat-kernel asymptotics for the Laplacian in the
presence of Dirichlet, or Robin, or mixed boundary conditions;
completely gauge-invariant boundary conditions in Euclidean
quantum gravity; local vs. non-local boundary-value problems in
one-loop Euclidean quantum theory via path integrals.
\vskip 10cm
\noindent
\centerline {\bf 1. Introduction}
\vskip 0.3cm
The aim of theoretical physics is to
provide a clear conceptual framework for the wide variety of natural 
phenomena, so that not only are we able to make accurate predictions
to be checked against observations, but the underlying mathematical
structures of the world we live in can be thoroughly
understood by the scientific community. What are therefore the
key elements of a mathematical description of the physical world?
Can we derive all basic equations of theoretical physics from a few
symmetry principles? What do they tell us about the origin and
evolution of the universe? Why is gravitation so peculiar with
respect to all other fundamental interactions?

The above questions have received careful consideration over the last
decades, and have led, in particular, to several approaches [1] to a
theory aiming at achieving a synthesis of quantum physics on the one
hand, and general relativity on the other hand. This remains, possibly, 
the most important task of theoretical physics. 
The need for a quantum theory of gravity is already suggested
from singularity theorems in classical cosmology [2]. Such theorems  
prove that the Einstein theory of general relativity leads to the
occurrence of space-time singularities in a generic way. 
At first sight one might be tempted to conclude
that a breakdown of all physical laws occurred in the past,
or that general relativity is severely incomplete, being unable to
predict what came out of a singularity. It has been therefore 
pointed out that all these pathological features result from the attempt
of using the Einstein theory well beyond its limit of validity, i.e.
at energy scales where the fundamental theory is definitely more 
involved. General relativity might be therefore viewed as a low-energy
limit of a richer theory [3], which achieves the synthesis of both the
basic principles of modern physics and the fundamental interactions
in the form presently known.  

Within the framework just outlined it remains however true that the
various approaches to quantum gravity developed so far suffer from
mathematical inconsistencies, or incompleteness in their ability
of accounting for some basic features of the laws of nature. From the
point of view of general principles, the space-time approach to 
quantum mechanics [4] and quantum field theory [5], and its application
to the quantization of gravitational interactions [6--9], remains indeed 
of fundamental importance. When one tries to implement the Feynman
sum over histories one discovers that, already at the level of
non-relativistic quantum mechanics, a well defined mathematical
formulation is only obtained upon considering a heat-equation
problem. The measure occurring in the Feynman representation of the
Green kernel is then meaningful, and the propagation amplitude of
quantum mechanics in flat Minkowski space-time is obtained by
analytic continuation. This is a clear indication that 
quantum-mechanical problems via path integrals are well understood 
only if the heat-equation counterpart is mathematically well posed.
In quantum field theory one then deals with the Euclidean approach,
and its application to quantum gravity [10] relies heavily on the theory
of elliptic operators on Riemannian 
manifolds [11--14]. To obtain a complete
picture one has then to specify the boundary conditions of the 
theory, i.e. the class of Riemannian geometries with their topologies
involved in the sum, and the form of boundary data assigned on
the bounding surfaces [15--29].

In particular, recent work [30,31] has shown
that the only set of local boundary conditions on metric
perturbations which are completely invariant under infinitesimal
diffeomorphisms is incompatible with a good elliptic
theory. More precisely, while the resulting operator on metric
perturbations can be made of Laplace type and elliptic in the
interior of the Riemannian manifold under consideration, the property
of strong ellipticity [12--14] is violated. This is a precise 
mathematical expression of the requirement that a unique smooth solution
of the boundary-value problem should exist which vanishes at infinite
geodesic distance from the boundary. This opens deep interpretive
issues, since only for gravity does the requirement of complete gauge
invariance of the boundary conditions turn out to be incompatible
with a good elliptic theory. It is then impossible to make sense
even just of the one-loop semiclassical approximation, because the
functional trace of the heat operator is found to diverge
(cf. section 6). 

We have been therefore led to consider, 
in our more recent research, non-local
boundary conditions for the quantized gravitational field at one-loop
level [32--35]. On the one hand, such a 
scheme already arises in simpler
problems, i.e. the quantum theory of a free particle subject to 
non-local boundary data on a circle [36]. One then finds two families of
eigenfunctions of the Hamiltonian: surface states which decrease
exponentially as one moves away from the boundary, and
bulk states which remain instead smooth and non-vanishing. 
The generalization to an Abelian gauge theory such as Maxwell theory
can fulfill non-locality, ellipticity and complete gauge invariance
of boundary conditions providing one learns to work with  
pseudo-differential operators in one-loop quantum theory [37]. On the
other hand, in the application to quantum gravity, since the boundary
operator acquires new kernels responsible for the pseudo-differential
nature of the boundary-value problem, one might hope to be able to
recover a good elliptic theory under a wider variety of 
conditions [32--35]. 

Boundary field theory
is therefore highly relevant for understanding quantum cosmology, 
quantum gravity and the foundations of quantized gauge theories, and it
has deep roots in global analysis and spectral geometry, since, after
imposing a supplementary condition, the gauge-field operator in the path
integral for such theories [5,9] is either of Laplace type or non-minimal
(on working with positive-definite metrics). In agreement 
with our pedagogical aims, we first review operators of Laplace
type in section 2, the functorial method for heat-kernel asymptotics
in section 3 and the topic of mixed boundary conditions in section 4.
Section 5 is then devoted to gauge-invariant boundary conditions
for the quantized gravitational field, while recent developments
and open problems are outlined in section 6.
\vskip 0.3cm
\centerline {\bf 2. Asymptotics of the Laplacian on manifolds with boundary}
\vskip 0.3cm
Following Branson and Gilkey [38], we are interested in a
se\-co\-nd-or\-der differential operator, say $P$, with leading
symbol given by the metric tensor (more precisely, with scalar
leading symbol) on a compact $m$-dimensional
Riemannian manifold $M$ with boundary $\partial M$. Denoting
by $\nabla$ the connection on the vector bundle $V$ over $M$,
our assumption implies that $P$, called an operator of
Laplace type, reads 
$$
P=-g^{ab}\nabla_{a}\nabla_{b}-E ,
\eqno (2.1)
$$
where $E$ is an endomorphism of $V$ (i.e. the `potential' term
in the physics-oriented literature). The heat equation for
the operator $P$ is 
$$
\left({\partial \over \partial t}+P \right)\varphi=0 .
\eqno (2.2)
$$
By definition, the {\it heat kernel} is the
solution, for $t > 0$, of the equation 
$$
\left({\partial \over \partial t}+P \right)
U(x,x';t)=0 ,
\; \; x \; {\rm and} \; x' \in \; M  ,
\eqno (2.3)
$$
subject to the boundary condition
$$
\Bigr[B \; U(x,x';t)\Bigr]_{\partial M}=0 ,
\eqno (2.4)
$$
jointly with the (initial) condition 
$$
\lim_{t \to 0^{+}} \int_{M}U(x,x';t) \rho(x')dx'
=\rho(x) ,
\eqno (2.5)
$$
which is a rigorous mathematical expression for the Dirac
delta behaviour as $t \rightarrow 0^{+}$.
The heat kernel can be written as 
$$
U(x,x';t)=\sum_{(n)}\varphi_{(n)}(x) \varphi_{(n)}(x')
e^{-\lambda_{(n)}t} ,
\eqno (2.6)
$$
where $\left \{ \varphi_{(n)}(x) \right \}$ is a complete
orthonormal set of eigenfunctions of $P$ with eigenvalues 
$\lambda_{(n)}$. The index $n$ is enclosed in round brackets,
to emphasize that, in general, a finite collection of
integer labels occurs therein. 

Since, by construction, the heat kernel behaves as a
distribution in the neighbourhood of the boundary, it is
convenient to introduce a smooth function, say 
$f \in C^{\infty}(M)$, and consider a slight generalization
of the trace function (or integrated heat kernel), 
i.e. ${\rm Tr}_{L^{2}} \Bigr(f e^{-tP} \Bigr)$. It is
precisely the consideration of $f$ that makes it possible to
deal properly with the distributional behaviour of the heat kernel
near $\partial M$. A key idea is therefore to work with
arbitrary $f$, and then set $f=1$ only when all coefficients
in the asymptotic expansion
$$ \eqalignno{
{\rm Tr}_{L^{2}}\Bigr(f e^{-t P} \Bigr) & \equiv
\int_{M}{\rm Tr}\Bigr[f U(x,x;t)\Bigr] \cr
&\sim (4\pi t)^{-m/2} \sum_{n=0}^{\infty}
t^{n/2} a_{n/2}(f,P,{\cal B})
&(2.7)\cr}
$$
have been evaluated. The term $U(x,x;t)$ is called the
{\it heat-kernel diagonal}. By virtue of Greiner's result [39],
the heat-kernel coefficients $a_{n/2}(f,P,{\cal B})$, 
which are said to describe the
{\it global asymptotics}, are obtained by integrating {\it local}
formulae. More precisely, they admit a split into integrals
over $M$ (interior terms) and over $\partial M$ (boundary
terms). In such formulae, the integrands are linear combinations
of all geometric invariants of the appropriate dimension
(see below)
which result from the Riemann curvature $R_{\; \; bcd}^{a}$
of the background, the extrinsic curvature of the boundary,
the differential operator $P$ (through the endomorphism 
$E$), and the boundary operator $\cal B$ (through the
endomorphisms, or projection operators, or more general
matrices occurring in it). With our notation, the indices
$a,b,...$ range from 1 through $m$ and index a local
orthonormal frame for the tangent bundle of $M$, $TM$, while
the indices $i,j,...$ range from 1 through $m-1$ and index
the orthonormal frame for the tangent bundle of the 
boundary, $T(\partial M)$. The boundary is defined by the
equations
$$
\partial M: \; \; \; \; \; \; \; \; 
y^{a}=y^{a}(x) ,
\eqno (2.8)
$$
in terms of the functions $y^{a}(x)$, $x^{i}$ being the
coordinates on $\partial M$, and the $y^{a}$ those on $M$.
Thus, the {\it intrinsic} metric, $\gamma_{ij}$, on the
boundary hypersurface $\partial M$, is given in terms of the
metric $g_{ab}$ on $M$ by 
$$
\gamma_{ij}=g_{ab} \; y_{\; ,i}^{a} \; y_{\; ,j}^{b} .
\eqno (2.9)
$$
On inverting this equation one finds (here, $n^{a}=N^{a}$
is the inward-pointing normal)
$$
g^{ab}=q^{ab}+n^{a}n^{b} ,
\eqno (2.10)
$$
where 
$$
q^{ab}=y_{\; ,i}^{a} \; y_{\; ,j}^{b} \; \gamma^{ij} .
\eqno (2.11)
$$
The tensor $q^{ab}$ is equivalent to $\gamma^{ij}$ and
may be viewed as the {\it induced metric} on 
$\partial M$, in its contravariant form. The tensor 
$q_{\; \; b}^{a}$ is a projection operator, in that
$$
q_{\; \; b}^{a} \; q_{\; \; c}^{b}=q_{\; \; c}^{a} ,
\eqno (2.12)
$$
$$
q_{\; \; b}^{a} \; n^{b}=0 .
\eqno (2.13)
$$
The extrinsic-curvature tensor $K_{ab}$ (or second fundamental
form of $\partial M$) is here defined by the projection of the
covariant derivative of an {\it extension} of the inward,
normal vector field $n$:
$$
K_{ab} \equiv n_{c;d} \; q_{\; \; a}^{c} 
\; q_{\; \; b}^{d} ,
\eqno (2.14)
$$
and is symmetric if the metric-compatible connection on
$M$ is torsion-free. Only its spatial components, $K_{ij}$,
are non-vanishing. 

The semicolon $;$ denotes multiple covariant differentiation
with respect to the Levi--Civita connection $\nabla_{M}$ 
of $M$, while the stroke $\mid$ denotes multiple covariant
differentiation tangentially with respect to the 
Levi--Civita connection $\nabla_{\partial M}$  
of the boundary. When sections of bundles built from $V$
are involved, the semicolon means
$$
\nabla_{M} \otimes \II + \II \otimes \nabla ,
$$
and the stroke means 
$$
\nabla_{\partial M} \otimes \II
+ \II \otimes \nabla .
$$
The curvature of the connection $\nabla$ on $V$ is denoted
by $\Omega$.

When Dirichlet or Robin boundary conditions are imposed on
sections of $V$:
$$
[\phi]_{\partial M}=0 ,
\eqno (2.15)
$$
or
$$
\Bigr[(n^{a}\nabla_{a}+S)\phi \Bigr]_{\partial M}=0 ,
\eqno (2.16)
$$
the global asymptotics in (2.7) is expressed through some
{\it universal constants}  
$$
\left \{ \alpha_{i},b_{i},
c_{i}, d_{i}, e_{i} \right \}
$$ 
such that (here 
$R_{ab} \equiv R_{\; \; abc}^{c}$ is the Ricci tensor,
$R \equiv R_{\; \; a}^{a}$, and $\cstok{\ } \equiv
\nabla^{a}\nabla_{a}=g^{ab}\nabla_{a}\nabla_{b}$)
$$
a_{0}(f,P,{\cal B})=\int_{M}{\rm Tr}(f) ,
\eqno (2.17)
$$
$$
a_{1/2}(f,P,{\cal B})=\delta (4\pi)^{1/2}
\int_{\partial M}{\rm Tr}(f) ,
\eqno (2.18)
$$
$$
\eqalignno{
\; & a_{1}(f,P,{\cal B})
={1\over 6}\int_{M}{\rm Tr}\Bigr[\alpha_{1}
fE+\alpha_{2}fR \Bigr] \cr
&+{1\over 6}\int_{\partial M}{\rm Tr} \Bigr[
b_{0}f({\rm tr}K)+b_{1}f_{;N}+b_{2}fS \Bigr] ,
&(2.19)\cr}
$$
$$
\eqalignno{
\; & a_{3/2}(f,P,{\cal B})
={\delta \over 96}(4\pi)^{1/2}
\int_{\partial M}{\rm Tr} \Bigr[f(c_{0}E+c_{1}R
+c_{2} R_{\; \; NiN}^{i} \cr
&+c_{3}({\rm tr}K)^{2}+c_{4}K_{ij}K^{ij}
+c_{7}S ({\rm tr}K)+c_{8}S^{2} \Bigr) \cr
&+f_{;N}\Bigr(c_{5}({\rm tr}K)+c_{9}S \Bigr)
+c_{6} f_{;NN} \Bigr] ,
&(2.20)\cr}
$$
$$
\eqalignno{
\; & a_{2}(f,P,{\cal B})={1\over 360} \int_{M}{\rm Tr}
\biggr[f \Bigr(\alpha_{3} \cstok{\ }E +\alpha_{4}RE
+\alpha_{5} E^{2} \cr
& + \alpha_{6} \cstok{\ }R 
+\alpha_{7}R^{2}
+\alpha_{8} R_{ab}R^{ab} \cr
&+\alpha_{9}R_{abcd}R^{abcd}
+\alpha_{10} \Omega_{ab} \Omega^{ab} \Bigr) \biggr]\cr
&+{1\over 360} \int_{\partial M}{\rm Tr} \biggr[
f \Bigr(d_{1} E_{;N}+d_{2}R_{;N}
+d_{3} ({\rm tr}K)_{\mid i}^{\; \; \; \mid i}
+d_{4} K_{ij}^{\; \; \; \mid ij} \cr
&+d_{5} E ({\rm tr}K)+d_{6}R ({\rm tr}K)
+d_{7} R_{\; \; NiN}^{i} ({\rm tr} K)
+d_{8} R_{iNjN} K^{ij} \cr
&+d_{9} R_{\; ilj}^{l} K^{ij}
+d_{10}({\rm tr}K)^{3}+d_{11}K_{ij}K^{ij}({\rm tr}K) \cr
&+d_{12}K_{i}^{\; \; j} \; K_{j}^{\; \; l} 
\; K_{l}^{\; \; i}
+d_{13} \Omega_{iN;}^{\; \; \; \; \; i}
+d_{14} SE \cr 
& + d_{15} SR +d_{16} S R_{\; \; NiN}^{i}
+d_{17} S ({\rm tr}K)^{2}+d_{18}S K_{ij} K^{ij} \cr
&+d_{19} S^{2} ({\rm tr}K)+d_{20} S^{3}
+d_{21} S_{\mid i}^{\; \; \; \mid i} \Bigr)
+f_{;N} \Bigr(e_{1}E + e_{2} R \cr
&+e_{3} R_{\; \; NiN}^{i}+e_{4}({\rm tr}K)^{2}
+e_{5}K_{ij}K^{ij}+e_{8}S ({\rm tr}K)+e_{9}S^{2}\Bigr)\cr
&+f_{;NN} \Bigr(e_{6} ({\rm tr}K)+e_{10}S \Bigr)
+e_{7} f_{;a \; \; N}^{\; \; \; a} \biggr] .
&(2.21)\cr}
$$
These formulae may seem to be very complicated, but there
is indeed a systematic way to write them down and then
compute the universal constants. To begin, note that,
if $k$ is odd, $a_{k/2}(f,P,{\cal B})$ 
receives contributions from
boundary terms only, whereas both interior terms and 
boundary terms contribute to 
$a_{k/2}(f,P,{\cal B})$, if $k$ is
even and positive. In the $a_{1}$ coefficient, the 
integrand in the interior term must be linear in the 
curvature, and hence it can only be a linear combination
of the trace of the Ricci tensor, and of the endomorphism
$E$ in the differential operator. In the $a_{2}$ 
coefficient, the integrand in the interior term must be
quadratic in the curvature, and hence one needs a linear
combination of the eight geometric invariants [6]
$$
\cstok{\ }E \; , \; 
RE \; , \; E^{2} \; , \;
\cstok{\ }R \; , \;
R^{2} \; , \; R_{ab}R^{ab} \; , \;
R_{abcd}R^{abcd} \; , \; \Omega_{ab}\Omega^{ab} .
$$

In the $a_{1}$ coefficient, the integrand in the boundary
term is a local expression given by a linear combination
of all invariants linear (or of the same dimension as terms
linear) in the extrinsic curvature:
${\rm tr}K, S$ and $f_{;N}$. In the $a_{3/2}$ coefficient,
the integrand in the boundary term must be quadratic in
the extrinsic curvature. Thus, bearing in mind the
Gauss--Codazzi equations, one finds the general result
(2.20). Last, in the $a_{2}$ coefficient, the integrand
in the boundary term must be cubic in the extrinsic 
curvature. This leads to the boundary integral in (2.21),
bearing in mind that $f_{;N}$ is 
on the same ground of a term linear in $K_{ij}$,
while $f_{;NN}$ is on the same ground of a
term quadratic in $K_{ij}$.
Note that the interior invariants are built universally and
polynomially from the metric tensor, its inverse, and the
covariant derivatives of $R_{\; bcd}^{a}, \Omega_{ab}$
and $E$. By virtue of Weyl's work on the invariants of the
orthogonal group [12,38], 
these polynomials can be formed using only
tensor products and contraction of tensor arguments. Here,
the structure group is $O(m)$. However, when a boundary occurs,
the boundary structure group is $O(m-1)$. Weyl's theorem is
used again to construct invariants as in the previous 
equations [38].
\vskip 0.3cm
\centerline {\bf 3. Functorial method}
\vskip 0.3cm
Let $T$ be a map which carries finite-dimensional vector
spaces into finite-dimensional vector spaces. Thus, to
every vector space $V$ one has an associated vector space
$T(V)$. The map $T$ is said to be a {\it continuous functor}
if, for all $V$ and $W$, the map
$$
T: Hom(V,W) \longrightarrow Hom (T(V),T(W))
$$
is continuous. 

In the theory of heat kernels, an application 
of the functorial method is 
the analysis of heat-equation asymptotics with respect to
conformal variations. Indeed, the behaviour of classical and
quantum field theories under conformal rescalings of the metric
$$
{\widehat g}_{ab}=\Omega^{2} \; g_{ab} ,
\eqno (3.1)
$$
with $\Omega$ a smooth function, is at the heart of many deep
properties: light-cone structure, conformal curvature (i.e.
the Weyl tensor), conformal-infinity techniques, massless
free-field equations, twistor equation, twistor spaces, 
Hodge-star operator in four dimensions, conformal anomalies
[26]. In the functorial method, one chooses $\Omega$ in the form
$$
\Omega=e^{\varepsilon f} ,
\eqno (3.2)
$$
where $\varepsilon$ is a real-valued parameter, and
$f \in C^{\infty}(M)$ is the smooth function considered
in section 2. One then deals with a one-parameter 
family of differential operators
$$
P(\varepsilon)=e^{-2 \varepsilon f} \; P(0) ,
\eqno (3.3)
$$
boundary operators
$$
{\cal B}(\varepsilon)=e^{-\varepsilon f} \; {\cal B}(0) ,
\eqno (3.4)
$$
connections $\nabla^{\varepsilon}$ on $V$,
endomorphisms $E(\varepsilon)$ of $V$, and metrics
$$
g_{ab}(\varepsilon)=e^{2 \varepsilon f} \; g_{ab}(0) .
\eqno (3.5)
$$
For example, the form (3.2) of the conformal factor should
be inserted into the general formulae which describe the
transformation of Christoffel symbols under conformal
rescalings:
$$
{\widehat \Gamma}_{\; bc}^{a}=\Gamma_{\; bc}^{a}
+ \Omega^{-1} \Bigr(\delta_{\; b}^{a} \; \Omega_{,c}
+\delta_{\; c}^{a} \; \Omega_{,b}
-g_{bc}g^{ad} \Omega_{,d} \Bigr) .
\eqno (3.6)
$$
This makes it possible to obtain the conformal-variation
formulae for the Riemann tensor $R_{\; bcd}^{a}$ and for
all tensors involving the effect of Christoffel symbols.
For the extrinsic-curvature tensor defined in Eq. (2.14)
one finds
$$
{\widehat K}_{ab}=\Omega K_{ab}-n_{a}\nabla_{b}\Omega
+g_{ab} \nabla_{(n)}\Omega ,
\eqno (3.7)
$$
which implies 
$$
K_{ij}(\varepsilon)=e^{\varepsilon f} \; K_{ij}(0)
-\varepsilon \; g_{ij} \; e^{\varepsilon f} \; f_{;N} .
\eqno (3.8)
$$

The application of these methods to heat-kernel asymptotics
relies on the work by Branson
and Gilkey [38]. Within this framework, a crucial role is
played by the following `functorial' formulae
($F$ being another smooth function):
$$
\left[{d\over d\varepsilon}a_{n/2}\Bigr(1,e^{-2\varepsilon f}
P(0)\Bigr)\right]_{\varepsilon=0}=(m-n)a_{n/2}(f,P(0)) ,
\eqno (3.9)
$$
$$
\left[{d\over d\varepsilon}a_{n/2}\Bigr(1,P(0)
-\varepsilon F \Bigr)\right]_{\varepsilon=0}
=a_{{n/2}-1}(F,P(0)) ,
\eqno (3.10)
$$
$$
\left[{d\over d\varepsilon}a_{n/2}\Bigr(e^{-2 \varepsilon f}F,
e^{-2 \varepsilon f}P(0)\Bigr)\right]_{\varepsilon=0}=0 .
\eqno (3.11)
$$
Equation (3.11) is obtained when $m=n+2$. These properties
can be proved by (formal) differentiation, as follows.

If the conformal variation of an operator of Laplace type
reads
$$
P(\varepsilon)=e^{-2 \varepsilon f} P(0)
-\varepsilon F ,
\eqno (3.12)
$$
one finds
$$
\eqalignno{
\; & \left[{d\over d\varepsilon} {\rm Tr}_{L^{2}}
\Bigr(e^{-t P(\varepsilon)} \Bigr)\right]_{\varepsilon=0}
={\rm Tr}_{L^{2}} \biggr[\Bigr(2tf P(0)+tF \Bigr)
e^{-t P(0)} \biggr] \cr
&=-2t {\partial \over \partial t}{\rm Tr}_{L^{2}}
\biggr(f \; e^{-t P(0)} \biggr)+t {\rm Tr}_{L^{2}}
\biggr(F \; e^{-t P(0)} \biggr) .
&(3.13)\cr}
$$
Moreover, by virtue of the asymptotic expansion (2.7), 
one has (the numerical factors $(4\pi)^{-m/2}$ are omitted
for simplicity, since they do not affect the form of Eqs.
(3.9)--(3.11); following Branson and Gilkey [38] one 
can, instead, absorb such factors into the definition of 
the coefficients $a_{n/2}(f,P)$)
$$
{\partial \over \partial t}{\rm Tr}_{L^{2}}
\biggr(f \; e^{-t P(0)} \biggr) \sim -{1\over 2}
\sum_{n=0}^{\infty}(m-n)t^{{n\over 2}-{m\over 2}-1}
\; a_{n/2}(f,P(0)) .
\eqno (3.14)
$$
Thus, if $F$ vanishes, Eqs. (3.13) and (3.14) lead 
to the result (3.9). By contrast, if $f$ is set to
zero, one has $P(\varepsilon)=P(0)-\varepsilon F$, which
implies
$$
\eqalignno{
\; & \left[{d\over d \varepsilon}{\rm Tr}_{L^{2}}
\Bigr(e^{-t P(\varepsilon)}\Bigr)\right]_{\varepsilon=0}
\sim t^{-m/2} \sum_{n=0}^{\infty}t^{{n/2}+1}
\; a_{n/2}(F,P(0)) \cr
&= t^{-m/2} \sum_{l=2}^{\infty}t^{l/2} \;
a_{{l/2}-1}(F,P(0)) ,
&(3.15)\cr}
$$
which leads in turn to Eq. (3.10). Last, to obtain the
result (3.11), one considers the two-parameter 
conformal variation
$$
P(\varepsilon,\gamma)=e^{-2 \varepsilon f}P(0)
-\gamma e^{-2 \varepsilon f} F .
\eqno (3.16)
$$
Now in Eq. (3.9) we first replace $n$ by $n+2$, and then
set $m=n+2$. One then has, from Eq. (3.16):
$$
{\partial \over \partial \varepsilon}a_{{n/2}+1}
(1,P(\varepsilon,\gamma))=0 .
\eqno (3.17)
$$
Equation (3.17) can be differentiated with respect to
$\gamma$, i.e. (see (3.10))
$$ 
\eqalignno{
0&= {\partial^{2}\over \partial \gamma \partial \varepsilon}
a_{{n/2}+1}(1,P(\varepsilon,\gamma))
={\partial \over \partial \varepsilon}
{\partial \over \partial \gamma}a_{{n/2}+1}
\Bigr(1,e^{-2 \varepsilon f}(P(0)-\gamma F)\Bigr) \cr
&={\partial \over \partial \varepsilon}a_{{n/2}}
\Bigr(e^{-2 \varepsilon f}F, e^{-2 \varepsilon f}P(0)\Bigr) ,
&(3.18)\cr}
$$
and hence Eq. (3.11) is proved.

To deal with Robin boundary conditions 
one needs another lemma, which is
proved following again Branson and Gilkey. What we obtain
is indeed a particular case of a more general property, which
is proved in section 6 of Ref. [28]. Our
starting point is $M$, a compact, connected one-dimensional
Riemannian manifold with boundary. In other words, one deals
with the circle or with a closed interval. If
$$
b: C^{\infty}(M) \longrightarrow {\Re}
$$
is a smooth, real-valued function, one can form the
first-order operator
$$
A \equiv {d\over dx}-b ,
\eqno (3.19)
$$
and its (formal) adjoint
$$
A^{\dagger} \equiv -{d\over dx}-b .
\eqno (3.20)
$$
From these operators, one can form the second-order operators
$$
D_{1} \equiv A^{\dagger} A = -\left[{d^{2}\over dx^{2}}
-b_{x}-b^{2}\right] ,
\eqno (3.21)
$$
$$
D_{2} \equiv A A^{\dagger}=-\left[{d^{2}\over dx^{2}}
+b_{x}-b^{2} \right] ,
\eqno (3.22)
$$
where $b_{x} \equiv {db\over dx}$. For $D_{1}$, Dirichlet
boundary conditions are taken, while Robin boundary conditions
are assumed for $D_{2}$. On defining
$$
f_{x} \equiv {df\over dx} , \;
f_{xx} \equiv {d^{2}f \over dx^{2}} ,
$$
one then finds the result 
$$
(n-1)\biggr[a_{n/2}(f,D_{1})-a_{n/2}(f,D_{2})\biggr]
=a_{{n/2}-1} \Bigr(f_{xx}+2b f_{x}, D_{1} \Bigr) .
\eqno (3.23)
$$
As a first step in the proof of (3.23), one takes a spectral
resolution for $D_{1}$, say $\left \{ \theta_{\nu}, \lambda_{\nu}
\right \}$, where $\theta_{\nu}$ is the eigenfunction of $D_{1}$
belonging to the eigenvalue $\lambda_{\nu}$:
$$
D_{1} \; \theta_{\nu}=\lambda_{\nu} \; \theta_{\nu} .
\eqno (3.24)
$$
Thus, differentiation with respect to $t$ of the
heat-kernel diagonal:
$$
U(D_{1},x,x;t)=\sum_{\nu}e^{-t \lambda_{\nu}}
\; \theta_{\nu}^{2}(x) ,
\eqno (3.25)
$$
yields
$$
{\partial \over \partial t}U(D_{1},x,x;t)
=-\sum_{\nu}\lambda_{\nu} e^{-t \lambda_{\nu}}
\; \theta_{\nu}^{2}(x)=-\sum_{\nu}e^{-t \lambda_{\nu}}
(D_{1} \; \theta_{\nu})\theta_{\nu} .
\eqno (3.26)
$$
Moreover, for any $\lambda_{\nu} \not = 0$, the set
$$
\left \{ {A \theta_{\nu}\over \sqrt{\lambda_{\nu}}},
\lambda_{\nu} \right \}
$$
provides a spectral resolution of $D_{2}$ on
${\rm Ker}(D_{2})^{\perp}$, and one finds
$$
\eqalignno{
\; & {\partial \over \partial t}U(D_{2},x,x;t)
=-\sum_{\lambda_{\nu} \not = 0} \lambda_{\nu}
e^{-t \lambda_{\nu}} \; \theta_{\nu}^{2}(x) \cr
&= -\sum_{\lambda_{\nu} \not = 0} e^{-t \lambda_{\nu}}
\Bigr(\sqrt{\lambda_{\nu}} \; \theta_{\nu} \Bigr)
\Bigr(\sqrt{\lambda_{\nu}} \; \theta_{\nu} \Bigr) \cr
&=-\sum_{\lambda_{\nu} \not = 0} e^{-t \lambda_{\nu}}
(A \theta_{\nu}) (A \theta_{\nu}) .
&(3.27)\cr}
$$
Bearing in mind that $A \theta_{\nu}=0$ if $\lambda_{\nu}=0$,
summation may be performed over all values of $\nu$, to find
$$
\eqalignno{
\; & 2 {\partial \over \partial t}\Bigr[U(D_{1},x,x;t)
-U(D_{2},x,x;t)\Bigr] \cr
&=2 \sum_{\nu}e^{-t \lambda_{\nu}}\biggr[\Bigr(
\theta_{\nu}'' \theta_{\nu}-b' \theta_{\nu}^{2}
-b^{2} \theta_{\nu}^{2}\Bigr)+(\theta_{\nu}'-b \theta_{\nu})
(\theta_{\nu}'-b \theta_{\nu})\biggr] \cr
&= 2 \sum_{\nu}e^{-t \lambda_{\nu}}\biggr[\theta_{\nu}''
\theta_{\nu}-b' \theta_{\nu}^{2}+(\theta_{\nu}')^{2}
-2b \theta_{\nu}' \theta_{\nu} \biggr] .
&(3.28)\cr}
$$
On the other hand, differentiation with respect to $x$ yields
$$
\left({\partial \over \partial x}-2b \right)U(D_{1},x,x;t)
=\sum_{\nu}e^{-t \lambda_{\nu}}\Bigr(2 \theta_{\nu}
\theta_{\nu}' -2b \theta_{\nu}^{2} \Bigr) ,
\eqno (3.29)
$$
which implies
$$ \eqalignno{
\; & f {\partial \over \partial x} \left({\partial \over \partial x}
-2b \right)U(D_{1},x,x;t) \cr
&=2f {\partial \over \partial t}
\Bigr[U(D_{1},x,x;t)-U(D_{2},x,x;t)\Bigr] .
&(3.30)\cr}
$$
We now integrate this formula over $M$ and use the boundary
conditions described before, jointly with integration by parts.
All boundary terms are found to vanish, so that
$$
\eqalignno{
\; & \int_{M}f {\partial \over \partial x}\left(
{\partial \over \partial x}-2b \right)U(D_{1},x,x;t)dx \cr
&=\int_{M} \left({\partial^{2}f \over \partial x^{2}}
+2b {\partial f \over \partial x}\right)
U(D_{1},x,x;t)dx \cr
&=\int_{M}2f {\partial \over \partial t} \Bigr[
U(D_{1},x,x;t)-U(D_{2},x,x;t)\Bigr] .
&(3.31)\cr}
$$
Bearing in mind the standard notation for heat-kernel
traces, Eq. (3.31) may be re-expressed as
$$
\eqalignno{
\; & 2 {\partial \over \partial t}\left[
{\rm Tr}_{L^{2}}\Bigr(f e^{-t D_{1}} \Bigr)
-{\rm Tr}_{L^{2}} \Bigr(f e^{-t D_{2}} \Bigr)\right] \cr
&={\rm Tr}_{L^{2}} \left[\Bigr(f_{xx}+2b f_{x}\Bigr)
e^{-t D_{1}} \right] ,
&(3.32)\cr}
$$
which leads to Eq. (3.23) by virtue of the asymptotic
expansion (2.7). 

The algorithm resulting from Eq. (3.10) is sufficient to
determine almost all interior terms in heat-kernel asymptotics.
To appreciate this, notice that one is dealing with conformal
variations which only affect the endomorphism of the operator
$P$ in (2.1). For example, on setting $n=2$ in Eq. (3.10),
one ends up by studying (the tilde symbol is now used for
interior terms)
$$
{\tilde a}_{1}(1,P) \equiv {1\over 6} \int_{M}{\rm Tr}
\Bigr[\alpha_{1}E+\alpha_{2}R \Bigr]
={\tilde a}_{1}(E,R) ,
\eqno (3.33)
$$
which implies
$$
\eqalignno{
\; & {\tilde a}_{1}(1,P(0)-\varepsilon F)
={\tilde a}_{1}(E,R)-{\tilde a}_{1}(E-\varepsilon F,R) \cr
&={1\over 6} \int_{M}{\rm Tr} \biggr[\alpha_{1}
(E-(E-\varepsilon F))+\alpha_{2}(R-R)\biggr] \cr
&={1\over 6} \int_{M}{\rm Tr}(\alpha_{1}\varepsilon F) ,
&(3.34)\cr}
$$
and hence
$$
\left[{d\over d \varepsilon}{\tilde a}_{1}
(1,P(0)-\varepsilon F)\right]_{\varepsilon=0}
={1\over 6} \int_{M} {\rm Tr}(\alpha_{1}F) 
={\tilde a}_{0}(F,P(0))=\int_{M}{\rm Tr}(F) .
\eqno (3.35) 
$$
By comparison, Eq. (3.35) shows that 
$$
\alpha_{1}=6 .
\eqno (3.36)
$$
An analogous procedure leads to (see (2.21))
$$
\eqalignno{
\; & {\tilde a}_{2}(1,P(0)-\varepsilon F)
={\tilde a}_{2}\Bigr(E,R,{\rm Ric},{\rm Riem},\Omega \Bigr) \cr
&-{\tilde a}_{2}\Bigr(E-\varepsilon F,R,{\rm Ric},{\rm Riem},
\Omega \Bigr) \cr
&={1\over 360} \int_{M}{\rm Tr} \Bigr[\alpha_{4}R \varepsilon F
+ \alpha_{5} (E^{2}-(E-\varepsilon F)^{2})\Bigr] \cr
&={1\over 360} \int_{M}{\rm Tr} \biggr[\alpha_{4}R \varepsilon
F + \alpha_{5}\Bigr(-{\varepsilon}^{2}F^{2}
+ 2 \varepsilon E F \Bigr)\biggr] .
&(3.37)\cr}
$$
Now one can apply Eq. (3.10) when $n=4$, to find
$$
\eqalignno{
\; & \left[{d\over d \varepsilon}{\tilde a}_{2}
(1,P(0)-\varepsilon F) \right]_{\varepsilon=0}
={1\over 360} \int_{M}{\rm Tr}\Bigr[\alpha_{4}FR
+2 \alpha_{5} FE \Bigr] \cr
&={\tilde a}_{1}(F,P(0))={1\over 6}\int_{M}{\rm Tr}
\Bigr[\alpha_{1}FE+\alpha_{2}FR \Bigr] .
&(3.38)\cr}
$$
Equating the coefficients of the invariants occurring
in the equation (3.38) one finds
$$
\alpha_{2}={1\over 60}\alpha_{4} ,
\eqno (3.39)
$$
$$
\alpha_{5}=30 \alpha_{1}=180 .
\eqno (3.40)
$$
Furthermore, the consideration of Eq. (3.11) when $n=2$ yields
$$
\eqalignno{
\; & \left[{d\over d\varepsilon}{\tilde a}_{1}
\Bigr(e^{-2 \varepsilon f}F, e^{-2 \varepsilon f} P(0)
\Bigr)\right]_{\varepsilon=0}={1\over 6}\int_{M}
\left \{ \biggr[{d\over d\varepsilon} {\rm Tr}
(\alpha_{1}FE)\biggr]_{\varepsilon=0} \right . \cr
&+2f {\rm Tr}(\alpha_{1}FE) \cr
& \left . + \left[{d\over d\varepsilon} {\rm Tr}
(\alpha_{2}FR)\right]_{\varepsilon=0}
+ 2f {\rm Tr}(\alpha_{2}FR) \right \} .
&(3.41)\cr}
$$
At this stage, we need the conformal-variation formulae
$$
\left[{d\over d\varepsilon}E(\varepsilon) 
\right]_{\varepsilon=0}
=-2fE+{1\over 2}(m-2) \cstok{\ } f ,
\eqno (3.42)
$$
$$
\left[{d\over d \varepsilon}R(\varepsilon) 
\right]_{\varepsilon=0}
=-2fR-2(m-1)\cstok{\ }f .
\eqno (3.43)
$$
Since we are studying the case $m=n+2=4$, we find
$$
\left[{d\over d \varepsilon}{\tilde a}_{1}
\Bigr(e^{-2 \varepsilon f}F, e^{-2 \varepsilon f}P(0)
\Bigr)\right]_{\varepsilon=0}={1\over 6} \int_{M}
{\rm Tr} \Bigr[(\alpha_{1}-6 \alpha_{2})F 
\cstok{\ }f \Bigr]=0 ,
\eqno (3.44)
$$
which implies
$$
\alpha_{2}={1\over 6} \alpha_{1}=1 ,
\eqno (3.45)
$$
$$
\alpha_{4}=60 \alpha_{2}=60 .
\eqno (3.46)
$$
After considering the Laplacian acting on functions for a
product manifold $M=M_{1} \times M_{2}$ (this is another
application of functorial methods), the complete set
of coefficients for interior terms can be determined [38]:
$$
\alpha_{3}=60 \; , \; \alpha_{6}=12 \; , \;
\alpha_{7}=5 \; , \; \alpha_{8}=-2 \; , \;
\alpha_{9}=2 \; , \; \alpha_{10}=30 .
\eqno (3.47)
$$

As far as interior terms are concerned, one has to use
Eqs. (3.9) and (3.11), jointly with two conformal-variation
formulae which provide divergence terms that
play an important role [38]: 
$$
\eqalignno{
\; & \left[{d\over d \varepsilon}a_{1} \Bigr(F,
e^{-2 \varepsilon f} P(0) \Bigr)\right]_{\varepsilon=0}
-(m-2) a_{1}(fF,P(0)) \cr
&={1\over 6}(m-4) \int_{M} {\rm Tr} \Bigr(F
\cstok{\ }f \Bigr) ,
&(3.48)\cr}
$$
$$
\eqalignno{
\; & \left[{d\over d \varepsilon}a_{2} \Bigr(F,
e^{-2 \varepsilon f}P(0) \Bigr)\right]_{\varepsilon=0}
-(m-4)a_{2} (fF,P(0)) \cr
&={1\over 360}(m-6) \int_{M} F {\rm Tr} \biggr[
6 f_{\; \; \; ;b}^{;b \; \; \; a}
+10 f^{;a} R \cr
&+60 f^{;a}E + 4 f_{;b} R^{ab} \biggr]_{;a} .
&(3.49)\cr}
$$
Equations (3.48) and (3.49) are proved for manifolds
without boundary in Lemma 4.2 of Ref. [38].  
They imply that, for manifolds with boundary, the right-hand
side of Eq. (3.48), evaluated for $F=1$, should be added
to the left-hand side of Eq. (3.9) when $n=2$. Similarly,
the right-hand side of Eq. (3.49), evaluated for $F=1$,
should be added to the left-hand side of Eq. (3.9)
when $n=4$. Other useful formulae involving boundary effects
are [38] 
$$
\int_{M} \Bigr(f_{;b} R^{ab} \Bigr)_{;a}
=\int_{\partial M} \biggr[f_{;j} \Bigr(
K_{\; \; \; \; \mid i}^{ij}-({\rm tr}K)^{\mid j}\Bigr)
+f_{;N} R_{\; \; NiN}^{i} \biggr] ,
\eqno (3.50)
$$
$$
f_{;i}^{\; \; \; ;i}=f_{\mid i}^{\; \; \; \mid i}
-({\rm tr}K)f_{;N} ,
\eqno (3.51)
$$
$$
\int_{M} \cstok{\ }f= - \int_{\partial M} f_{;N} ,
\eqno (3.52)
$$
$$
\int_{\partial M} \cstok{\ } f
=\int_{\partial M} \Bigr[f_{;NN}-({\rm tr}K)f_{;N}\Bigr] .
\eqno (3.53)
$$

On taking into account Eqs. (3.48)--(3.53), the application
of Eq. (3.9) when $n=2,3,4$ leads to 18 equations which are
obtained by setting to zero the coefficients multiplying
$$
f_{;N} \; \; ({\rm when} \; n=2) ,
$$
$$
f_{;NN} \; \; , \; \; f_{;N}({\rm tr}K) \; \; , \; \; 
f_{;N}S \; \; ({\rm when} \; n=3) ,
$$
and
$$
f_{;a \; \; \; N}^{\; \; ;a} \; \; , \; \;
f_{;N}E \; \; , \; \; 
f_{;N}R \; \; , \; \; 
f_{;N} R_{\; \; NiN}^{i} ,
$$
$$
f_{;NN} ({\rm tr}K) \; \; , \; \;
f_{;N}({\rm tr}K)^{2} \; \; , \; \;
f_{;N}K_{ij}K^{ij} \; \; , \; \; 
f_{\mid i}({\rm tr}K)^{\mid i} ,
$$
$$
f_{\mid i}K_{\; \; \; \mid j}^{ij} \; \; , \; \;
f_{\mid i} \Omega_{\; N}^{i} \; \; , \; \; 
f_{;N}S ({\rm tr}K) \; \; , \; \; 
f_{;N}S^{2} ,
$$
$$
f_{;NN}S \; \; , \; \; 
f_{\mid i}^{\; \; \; \mid i} \; S
\; \; ({\rm when} \; n=4) .
$$
The integrals of these 18 terms have a deep geometric nature
in that they form a basis for the integral invariants. The
resulting 18 equations are
$$
-b_{0}(m-1)-b_{1}(m-2)+{1\over 2}b_{2}(m-2)-(m-4)=0 ,
\eqno (3.54)
$$
$$
{1\over 2}c_{0}(m-2)-2c_{1}(m-1)+c_{2}(m-1)-c_{6}(m-3)=0 ,
\eqno (3.55)
$$
$$
\eqalignno{
\; & -{1\over 2}c_{0}(m-2)+2c_{1}(m-1)-c_{2}-2c_{3}(m-1) \cr
&-2c_{4}-c_{5}(m-3)+{1\over 2}c_{7}(m-2)=0 ,
&(3.56)\cr}
$$
$$
-c_{7}(m-1)+c_{8}(m-2)-c_{9}(m-3)=0 ,
\eqno (3.57)
$$
$$
-6(m-6)+{1\over 2}d_{1}(m-2)-2d_{2}(m-1)-e_{7}(m-4)=0 ,
\eqno (3.58)
$$
$$
-60(m-6)-2d_{1}-d_{5}(m-1)+{1\over 2}d_{14}(m-2)
-e_{1}(m-4)=0 ,
\eqno (3.59)
$$
$$
-10(m-6)-2d_{2}-d_{6}(m-1)
+d_{9}+{1\over 2}d_{15}(m-2)
-e_{2}(m-4)=0 ,
\eqno (3.60) 
$$
$$
-d_{7}(m-1)-d_{8}+2d_{9}-e_{3}(m-4) 
+{1\over 2}d_{16}(m-2)
+4(m-6)=0 ,
\eqno (3.61) 
$$
$$
{1\over 2}d_{5}(m-2)-2d_{6}(m-1)+d_{7}(m-1)+d_{8}
-e_{6}(m-4)=0 ,
\eqno (3.62)
$$
$$ 
\eqalignno{
\; & -{1\over 2}d_{5}(m-2)+2d_{6}(m-1)-d_{7}-d_{9}
-3d_{10}(m-1) \cr
&-2d_{11}+{1\over 2}d_{17}(m-2)-e_{4}(m-4)=0 ,
&(3.63)\cr}
$$
$$
\eqalignno{
\; & -d_{8}-d_{9}(m-3)-d_{11}(m-1)-3d_{12}\cr 
&+{1\over 2}
d_{18}(m-2)-e_{5}(m-4)=0 ,
&(3.64)\cr}
$$
$$
d_{3}(m-4)-{1\over 2}d_{5}(m-2)+2d_{6}(m-1)
-d_{7}-d_{9}-4(m-6)=0 ,
\eqno (3.65)
$$
$$
d_{4}(m-4)-d_{8}-d_{9}(m-3)+4(m-6)=0 ,
\eqno (3.66)
$$
$$
(m-4)d_{13}=0 ,
\eqno (3.67)
$$
$$ 
\eqalignno{
\; & -{1\over 2}d_{14}(m-2)+2d_{15}(m-1)-d_{16}
-2d_{17}(m-1) \cr
&-2d_{18}+d_{19}(m-2)-e_{8}(m-4)=0 ,
&(3.68)\cr}
$$
$$
-d_{19}(m-1)+{3\over 2}d_{20}(m-2)-e_{9}(m-4)=0 ,
\eqno (3.69)
$$
$$
{1\over 2}d_{14}(m-2)-2d_{15}(m-1)+d_{16}(m-1)
-e_{10}(m-4)=0 ,
\eqno (3.70)
$$
$$
{1\over 2}d_{14}(m-2)-2d_{15}(m-1)+d_{16}
-d_{21}(m-4)=0 .
\eqno (3.71)
$$
This set of algebraic equations among universal constants holds
independently of the choice of Dirichlet or Robin boundary
conditions. Another set of equations which hold for either 
Dirichlet or Robin boundary conditions is obtained by applying
Eq. (3.11) when $n=3,4$. One then obtains five equations which
result from setting to zero the coefficients multiplying
$$
F_{;N} f_{;N} \; \; ({\rm when} \; n=3) ,
$$
$$
f_{;N}F_{;NN} \; \; , \; \;
f_{;NN} F_{;N} \; \; , \; \; 
f_{;N} F_{;N} ({\rm tr}K) \; \; , \; \; 
f_{;N} F_{;N} S \; \; ({\rm when} \; n=4) .
$$
The explicit form of these equations is [38] 
$$
-4 c_{5}-5c_{6}+{3\over 2}c_{9}=0 ,
\eqno (3.72)
$$
$$
-5e_{6}-4e_{7}+2e_{10}=0 ,
\eqno (3.73)
$$
$$
2e_{1}-10e_{2}+5e_{3}-2e_{7}=0 ,
\eqno (3.74)
$$
$$
-2e_{1}+10e_{2}-e_{3}-10e_{4}-2e_{5}-5e_{6}+6e_{7}
+2e_{8}=0 ,
\eqno (3.75)
$$
$$
-5e_{8}+4e_{9}-5e_{10}=0 .
\eqno (3.76)
$$

Last, one has to use the Lemma expressed by Eq. (3.23)
when $n=2,3,4$, bearing in mind that
$$
E_{1} \equiv E(D_{1})=-b_{x}-b^{2} ,
\eqno (3.77)
$$
$$
E_{2} \equiv E(D_{2})=b_{x}-b^{2} .
\eqno (3.78)
$$
For example, when $n=2$, one finds
$$
a_{1}(f,D_{1})-a_{1}(f,D_{2})-a_{0}\Bigr(f_{xx}
+2bf_{x},D_{1} \Bigr)=0 ,
\eqno (3.79)
$$
which implies (with Dirichlet conditions for $D_{1}$ and Robin
conditions for $D_{2}$)
$$
\int_{M}\Bigr[6f(E_{1}-E_{2})-6f_{xx}-12S f_{x} \Bigr]
+\int_{\partial M}\Bigr[-b_{2}fS-(3+b_{1})f_{;N}\Bigr]=0 .
\eqno (3.80)
$$
The integrand of the interior term in Eq. (3.80) may be
re-expressed as a total divergence, and hence one gets
$$
\int_{\partial M} \Bigr[(12-b_{2})fS+(3-b_{1})f_{;N}\Bigr]=0 ,
\eqno (3.81)
$$
which leads to
$$
b_{1}=3 \; \; , \; \; b_{2}=12 .
\eqno (3.82)
$$
Further details concern only the repeated application of all
these algorithms, and hence we refer the reader to
Branson and Gilkey [38]. We should emphasize, however, that no
proof exists, so far, that functorial methods lead to the 
complete calculation of {\it all} heat-kernel coefficients.
For the time being one can only say that, when Dirichlet or
Robin boundary conditions, or a mixture of the two (see
section 4) are imposed, functorial methods have been
completely successful up to the evaluation of the $a_{5/2}$
coefficient (see Ref. [27] and references therein).
\vskip 0.3cm
\centerline {\bf 4. Mixed boundary conditions}
\vskip 0.3cm
Mixed boundary conditions are found to 
occur naturally in the theory of
fermionic fields, gauge fields and gravitation, in that some
components of the field obey one set of boundary conditions,
and the remaining part of the field obeys a complementary set
of boundary conditions [25].
Here, we focus on some mathematical
aspects of the problem. 
The framework of our
investigation consists, as in section 3, of a compact
Riemannian manifold, say $M$, with smooth boundary 
$\partial M$. Given a vector bundle $V$ over $M$, we assume
that $V$ can be decomposed as the direct sum
$$
V=V_{n} \oplus V_{d} ,
\eqno (4.1)
$$
near $\partial M$. The corresponding projection operators
are denoted by $\Pi_{n}$ and $\Pi_{d}$, respectively. On
$V_{n}$ one takes Neumann boundary conditions modified by some
endomorphism, say $S$, of $V_{n}$ (see (2.16)), while
Dirichlet boundary conditions hold on $V_{d}$. The (total)
boundary operator reads therefore [12] 
$$
{\cal B}f \equiv \biggr[\Bigr(\Pi_{n}f \Bigr)_{;N}
+S \Pi_{n}f \biggr]_{\partial M} \oplus
\Bigr[ \Pi_{d}f \Bigr]_{\partial M} .
\eqno (4.2)
$$
On defining
$$
\psi \equiv \Pi_{n}-\Pi_{d} ,
\eqno (4.3)
$$
seven new universal constants are found to contribute to
heat-kernel asymptotics for an operator of Laplace type,
say $P$. In other words, the linear combination of
projectors considered in Eq. (4.3) gives rise to seven
new invariants in the calculation of heat-kernel coefficients
up to $a_{2}$: one invariant contributes to 
${\tilde a}_{3/2}(f,P,{\cal B})$, whereas 
the other six contribute to 
${\tilde a}_{2}(f,P,{\cal B})$ (of course, the
number of invariants is continuously increasing as one 
considers higher-order heat-kernel coefficients). The 
dependence on the boundary operator is emphasized by 
including it explicitly into the arguments of heat-kernel
coefficients. One can thus write the general formulae
(cf. (2.20) and (2.21))
$$
{\tilde a}_{3/2}(f,P,{\cal B})={\delta \over 96}
(4\pi)^{1/2} \int_{\partial M}{\rm Tr} \Bigr[
\beta_{1} f \psi_{\mid i} \; \psi^{\mid i} \Bigr]
+a_{3/2}(f,P,{\cal B}) ,
\eqno (4.4)
$$
$$
\eqalignno{
{\tilde a}_{2}(f,P,{\cal B})&={1\over 360}
\int_{\partial M}{\rm Tr} \Bigr[\beta_{2}f \psi \psi_{\mid i}
\; \Omega_{\; N}^{i}+\beta_{3}f \psi_{\mid i} \; 
\psi^{\mid i} ({\rm tr}K) \cr
&+\beta_{4}f \psi_{\mid i} \; \psi_{\mid j} K^{ij}
+\beta_{5}f \psi_{\mid i} \; \psi^{\mid i} S \cr
&+ \beta_{6} f_{;N} \psi_{\mid i} \; \psi^{\mid i}
+\beta_{7} f \psi_{\mid i} \; \Omega_{\; N}^{i} \Bigr]
+a_{2}(f,P,{\cal B}) ,
&(4.5)\cr}
$$
where $a_{3/2}(f,P,{\cal B})$ is formally analogous to
Eq. (2.20), but with some universal constants replaced by
linear functions of $\psi, \Pi_{n}, \Pi_{d}$, and similarly
for $a_{2}(f,P,{\cal B})$ and Eq. (2.21). The work in Refs. [12,40]
has fixed the following values for the universal constants 
$\left \{ \beta_{i} \right \}$ occurring in Eqs. (4.4)
and (4.5):
$$
\beta_{1}=-12 ,  \; \beta_{2}=60 , \; 
\beta_{3}=-12 , \; \beta_{4}=-24 , \;
\beta_{5}=-120 , \; 
\beta_{6}=-18 , \; 
\beta_{7}=0 .
\eqno (4.6)
$$
To obtain this result, it is crucial to bear in mind that
the correct functorial formula for the endomorphism $S$ is
$$
\left[{d\over d \varepsilon}S(\varepsilon) 
\right]_{\varepsilon=0}
=-fS +{1\over 2}(m-2)f_{;N} \Pi_{n} .
\eqno (4.7)
$$
This result was first obtained in Ref. [40], where the
author pointed out that $\Pi_{n}$ should be included, since
the variation of $S$ should compensate another suitable variation
only on the subspace $V_{n}$. The unfortunate omission of
$\Pi_{n}$ led to incorrect results in physical applications, 
which were later corrected by Moss and Poletti [41], hence
confirming the analytic results in Refs. [18,21].
\vskip 5cm
\centerline {\bf 5. Gauge-invariant boundary conditions for the 
gravitational field}
\vskip 0.3cm
For gauge fields and gravitation, the boundary conditions
are mixed, in that some components of the field (more
precisely, a one-form or a two-form) obey a set of boundary
conditions, and the remaining part of the field obeys another
set of boundary conditions. Moreover, the boundary conditions 
are invariant under local gauge transformations provided that 
suitable boundary conditions are imposed on the corresponding
ghost zero-form or one-form.

We are here interested in the derivation of mixed boundary
conditions for Euclidean quantum gravity. The knowledge of
the classical variational problem, and the principle of
gauge invariance, are enough to lead to a highly non-trivial
boundary-value problem. Indeed, it is by now well-known
that, if one fixes the three-metric at the boundary in 
general relativity, the corresponding variational problem
is well-posed and leads to the Einstein equations, providing
the Einstein-Hilbert action is supplemented by a boundary
term whose integrand is proportional to the trace of the
second fundamental form [42]. In the corresponding
`quantum' boundary-value problem, which is relevant for the
one-loop approximation in quantum gravity, the perturbations
$h_{ij}$ of the induced three-metric are set to zero at the
boundary. Moreover, the whole set of metric perturbations
$h_{\mu \nu}$ are subject to the so-called infinitesimal
{\it gauge} transformations 
$$
{\widehat h}_{\mu \nu} \equiv h_{\mu \nu}
+\nabla_{(\mu} \; \varphi_{\nu)},
\eqno (5.1)
$$
where $\nabla$ is the Levi-Civita connection of the background
four-geometry with metric $g$, and $\varphi_{\nu}dx^{\nu}$ 
is the ghost one-form. In geometric language, the
difference between ${\widehat h}_{\mu \nu}$ and
$h_{\mu \nu}$ is given by the Lie derivative along $\varphi$
of the four-metric $g$. 

For problems with boundary, Eq. (5.1) implies that
$$
{\widehat h}_{ij}=h_{ij}+\varphi_{(i \mid j)}
+K_{ij} \varphi_{0},
\eqno (5.2)
$$
where the stroke denotes, as usual, three-dimensional 
covariant differentiation tangentially with respect to the
intrinsic Levi-Civita connection of the boundary, while
$K_{ij}$ is the extrinsic-curvature tensor of the boundary.
Of course, $\varphi_{0}$ and $\varphi_{i}$ are the normal
and tangential components of the ghost one-form, respectively.
Note that boundaries make it necessary to perform a 3+1
split of space-time geometry and physical fields.
As such, they introduce non-covariant elements in the analysis
of problems relevant for quantum gravity. This seems to be 
an unavoidable feature, although the boundary conditions may
be written in tensor language.

In the light of (5.2), the boundary conditions
$$
\Bigr[h_{ij}\Bigr]_{\partial M}=0
\eqno (5.3a)
$$
are gauge invariant, i.e.
$$
\Bigr[{\widehat h}_{ij}\Bigr]_{\partial M}=0,
\eqno (5.3b)
$$
if and only if the whole ghost one-form obeys homogeneous
Dirichlet conditions, so that
$$
\Bigr[\varphi_{0}\Bigr]_{\partial M}=0,
\eqno (5.4)
$$
$$
\Bigr[\varphi_{i}\Bigr]_{\partial M}=0.
\eqno (5.5)
$$
The conditions (5.4) and (5.5) are necessary and sufficient
since $\varphi_{0}$ and $\varphi_{i}$ are independent, and
three-dimensional covariant differentiation commutes with the
operation of restriction at the boundary. Indeed, we are 
assuming that the boundary is smooth and not totally geodesic,
i.e. $K_{ij} \not = 0$. However, at those points of $\partial M$
where the extrinsic-curvature tensor vanishes, the condition
(5.4) is no longer necessary.

The problem now arises to impose boundary conditions on the
remaining set of metric perturbations. The key point is to
make sure that {\it the invariance of such boundary conditions 
under the infinitesimal transformations (5.1) is again 
guaranteed by (5.4) and (5.5)},
since otherwise one would obtain incompatible sets of
boundary conditions on the ghost one-form. Indeed, on using
the Faddeev-Popov formalism for the amplitudes of quantum
gravity, it is necessary to use a gauge-averaging term in
the Euclidean action, of the form 
$$
I_{\rm g.a.} \equiv {1\over 32 \pi G \alpha}
\int_{M}\Phi_{\nu}\Phi^{\nu}\sqrt{{\rm det} \; g} \; d^{4}x,
\eqno (5.6)
$$
where $\Phi_{\nu}$ is any gauge-averaging 
functional which leads to self-adjoint elliptic operators
on metric and ghost perturbations. One then finds that
$$
\delta \Phi_{\mu}(h) \equiv 
\Phi_{\mu}(h)-\Phi_{\mu}(\widehat h)
={\cal F}_{\mu}^{\; \; \nu} \; \varphi_{\nu},
\eqno (5.7)
$$
where ${\cal F}_{\mu}^{\; \; \nu}$ is an elliptic operator
that acts linearly on the ghost one-form. Thus, if one
imposes the boundary conditions
$$
\Bigr[\Phi_{\mu}(h)\Bigr]_{\partial M}=0,
\eqno (5.8)
$$
and if one assumes that the ghost field can be expanded in a
complete orthonormal set of eigenfunctions $u_{\nu}^{(\lambda)}$
of ${\cal F}_{\mu}^{\; \; \nu}$ which vanish at the boundary, i.e.
$$
{\cal F}_{\mu}^{\; \; \nu} \; u_{\nu}^{(\lambda)}
=\lambda u_{\mu}^{(\lambda)},
\eqno (5.9)
$$
$$
\varphi_{\nu}=\sum_{\lambda}C_{\lambda}u_{\nu}^{(\lambda)},
\eqno (5.10)
$$
$$
\Bigr[u_{\mu}^{(\lambda)}\Bigr]_{\partial M}=0,
\eqno (5.11)
$$
the boundary conditions (5.8) are automatically gauge-invariant
under the Dirichlet conditions (5.4) and (5.5) on the ghost.

At a deeper level, the boundary conditions (5.3)--(5.5) and (5.8)
are invariant under Becchi--Rouet--Stora--Tyutin transformations
[43], but their consideration has been abandoned after the proof
of resulting lack of strong ellipticity in Refs. [30,31].
\vskip 0.3cm
\centerline {\bf 6. Recent developments and open problems}
\vskip 0.3cm
The boundary conditions involving tangential derivatives of the
field have been studied in detail not only in Refs. 
[25,28,30,31] but also in Refs. [44--46]. In particular, the
work in Refs. [45,46] has produced valuable results in 
heat-kernel asymptotics which are very important to get a 
geometric understanding of one-loop divergences in quantum 
field theory.

As far as Euclidean quantum gravity is concerned, we can see three
main alternatives if it is approached from the point of view 
of spectral geometry (for a modern perspective on yet other issues
the reader is referred to the paper by Vassilevich in this volume):
\vskip 0.3cm
\noindent
(i) If one insists on using the completely gauge-invariant boundary
conditions of section 5 when a supplementary (i.e. gauge-fixing)
condition of de Donder type is used [47], one should prove that,
since the $L^{2}$ trace of the heat semigroup can be then split into
a regular part and a singular part [31], a regularized $\zeta(0)$ 
value can be defined (cf. Ref. [48]) which is only affected by the
regular part of ${\rm Tr}_{L^{2}}(e^{-tP})$.
\vskip 0.3cm
\noindent
(ii) One can instead choose local and strongly elliptic boundary
conditions for quantum gravity, which are, however, only partially
gauge-invariant [49].
\vskip 0.3cm
\noindent
(iii) Last, but not least, one can resort to non-local field
theory by considering a non-local gauge-fixing functional in the
path integral. If one could prove that the resulting set of
symbols for the gauge-field operator and boundary operator which
are compatible with strong ellipticity of the boundary-value
problem [32--35] is non-empty, one would understand from first
principles why the universe starts in a quantum state but eventually
reaches a classical regime, since non-local boundary conditions
may lead to `surface states' having precisely this property [36].

Moreover, it remains to be seen whether such investigations have
an impact on current developments in string and brane theory [50].
\vskip 0.3cm
{\bf Acknowledgments}. The work of G. Esposito has been
partially supported by PRIN 2002 {\it SINTESI}. The author is 
indebted to Ivan Avramidi, Alexander Kamenshchik and Klaus 
Kirsten for scientific collaboration over many years, 
to Andrei Barvinsky, Ian Moss and Dmitri Vassilevich for 
correspondence, and to the Editors for scientific advice and
encouragement.
\vskip 0.3cm
\centerline {\bf References}
\vskip 0.3cm
\noindent
\item {1.}
G. Gibbons, S. Rankin and P. Shellard, {\it The future of
theoretical physics and cosmology}, Cambridge
University Press, 2003.
\item {2.}
S. W. Hawking and G. F. R. Ellis, {\it The large-scale 
structure of space-time}, Cambridge University Press, 1973.
\item {3.}
J. Polchinski, {\it String theory, vols. 1 and 2},
Cambridge University Press, 1998.
\item {4.}
R. P. Feynman, {\it The space-time approach to non-relativistic
quantum mechanics}, Rev. Mod. Phys. {\bf 20} (1948) pp.367--387. 
\item {5.}
B. S. DeWitt, {\it The space-time approach to quantum field theory},
pp.381--738, in: {\it Relativity, groups and topology II}, eds. 
B. S. DeWitt and R. Stora, North-Holland, 1984.
\item {6.}
B. S. DeWitt, {\it Dynamical theory of groups and fields},
Gordon and Breach, 1965.
\item {7.}
I. G. Avramidi, {\it Heat kernel and quantum gravity},
Springer--Verlag, 2000.
\item {8.}
G. Esposito, G. Miele and B. Preziosi, {\it Quantum gravity and
spectral geometry}, Nuclear Physics B Proceedings Supplement,
{\bf 104}, 2002.
\item {9.}
B. S. DeWitt, {\it The global approach to quantum field theory},
Oxford University Press, 2003.
\item {10.}
G. Gibbons and S. W. Hawking, {\it Euclidean quantum gravity},
World Scientific, 1993.
\item {11.}
B. Booss--Bavnbek and K. P. Wojciechowski, {\it Elliptic boundary
problems for Dirac operators}, Birkh\"{a}user, 1993.
\item {12.}
P. B. Gilkey, {\it Invariance theory, the heat equation and the 
Atiyah--Singer index theorem}, CRC Press, 1995.
\item {13.}
G. Grubb, {\it Functional calculus of pseudo-differential boundary
problems}, Birkh\"{a}user, 1996.
\item {14.}
K. Kirsten, {\it Spectral functions in mathematics and physics},
CRC Press, 2001.
\item {15.}
S. W. Hawking, {\it The boundary conditions of the universe},
Pont. Acad. Sci. Scri. Var. {\bf 48} (1982) pp.563--574.
\item {16.}
J. B. Hartle and S. W. Hawking, {\it Wave function of the universe},
Phys. Rev. D {\bf 28} (1983) pp.2960--2975.
\item {17.}
S. W. Hawking, {\it The boundary conditions for gauged supergravity},
Phys. Lett. B {\bf 126} (1983) pp.175--177.
\item {18.}
P. D. D'Eath and G. Esposito, {\it Local boundary conditions for
the Dirac Operator and one-loop quantum cosmology},
Phys. Rev. D {\bf 43} (1991) pp.3234--3248.
\item {19.}
P. D. D'Eath and G. Esposito, {\it Spectral boundary conditions in
one-loop quantum cosmology}, Phys. Rev. D {\bf 44} (1991)
pp.1713--1721.
\item {20.}
G. Esposito, {\it Gauge-averaging functionals for Euclidean Maxwell
theory in the presence of boundaries}, Class. Quantum Grav.
{\bf 11} (1994) pp.905--926.
\item {21.}
G. Esposito, A. Yu. Kamenshchik, I. V. Mishakov and G. Pollifrone,
{\it Euclidean Maxwell theory in the presence of boundaries. II},
Class. Quantum Grav. {\bf 11} (1994) pp.2939--2950.
\item {22.}
G. Esposito and A. Yu. Kamenshchik, {\it Mixed boundary conditions
in Euclidean quantum gravity}, Class. Quantum Grav. {\bf 12}
(1995) pp.2715--2722.
\item {23.}
G. Esposito and A. Yu. Kamenshchik, {\it One-loop divergences in
simple supergravity: boundary effects}, Phys. Rev. D {\bf 54}
(1996) pp.3869--3881.
\item {24.}
I. G. Avramidi, G. Esposito and A. Yu. Kamenshchik,
{\it Boundary operators in Euclidean quantum gravity}, Class. 
Quantum Grav. {\bf 13} (1996) pp.2361--2373.
\item {25.}
G. Esposito, A. Yu. Kamenshchik and G. Pollifrone,
{\it Euclidean quantum gravity on manifolds with boundary},
Kluwer, 1997.
\item {26.}
G. Esposito, {\it Dirac operators and spectral geometry},
Cambridge University Press, 1998 (hep-th 9704016).
\item {27.}
K. Kirsten, {\it The a5 coefficient on a manifold with boundary},
Class. Quantum Grav. {\bf 15} (1998) pp.L5--L12.
\item {28.}
I. G. Avramidi and G. Esposito, {\it New invariants in the
one-loop divergences on manifolds with boundary},
Class. Quantum Grav. {\bf 15} (1998) pp.281--297.
\item {29.}
G. Esposito and A. Yu. Kamenshchik, {\it Fourth-order operators
on manifolds with a boundary}, Class. Quantum Grav. 
{\bf 16} (1999) pp.1097--1111.
\item {30.}
I. G. Avramidi and G. Esposito, {\it Lack of strong ellipticity in
Euclidean quantum gravity}, Class. Quantum Grav.
{\bf 15} (1998) pp.1141--1152.
\item {31.}
I. G. Avramidi and G. Esposito, {\it Gauge theories on 
manifolds with boundary}, Commun. Math. Phys.
{\bf 200} (1999) pp.495--543.
\item {32.}
G. Esposito, {\it Non-local boundary conditions in Euclidean
quantum gravity}, Class. Quantum Grav. {\bf 16} (1999)
pp.1113--1126.
\item {33.}
G. Esposito, {\it New kernels in quantum gravity},
Class. Quantum Grav. {\bf 16} (1999) pp.3999--4010.
\item {34.}
G. Esposito, {\it Boundary operators in quantum field theory},
Int. J. Mod. Phys. A {\bf 15} (2000) pp.4539--4555.
\item {35.}
G. Esposito, {\it Quantum gravity in four dimensions},
Nova Science, 2001.
\item {36.}
M. Schr\"{o}der, {\it On the Laplace operator with non-local boundary
conditions and Bose condensation}, Rep. Math. Phys. {\bf 27}
(1989) pp.259--269.
\item {37.}
G. Esposito and C. Stornaiolo, {\it Non-locality and ellipticity in a
gauge-invariant quantization}, Int. J. Mod. Phys. A {\bf 15}
(2000) pp.449--460.
\item {38.}
T. P. Branson and P. B. Gilkey, {\it The asymptotics of the Laplacian
on a manifold with boundary}, Commun. Part. Diff. Eqs.
{\bf 15} (1990) pp.245--272.
\item {39.}
P. Greiner, {\it An asymptotic expansion for the heat equation},
Arch. Rat. Mech. Anal. {\bf 41} (1971) pp.163--218.
\item {40.}
D. V. Vassilevich, {\it Vector fields on a disk with mixed boundary
conditions}, J. Math. Phys. {\bf 36} (1995) pp.3174--3182.
\item {41.}
I. G. Moss and S. Poletti, {\it Conformal anomalies on Einstein spaces
with boundary}, Phys. Lett. B {\bf 333} (1994) pp.326--330.
\item {42.}
J. W. York Jr., {\it Boundary Terms in the Action Principles of
General Relativity}, Found. Phys. {\bf 16} (1986) pp.249--257.
\item {43.}
I. G. Moss and P. J. Silva, {\it BRST invariant boundary conditions for
gauge theories}, Phys. Rev. D {\bf 55} (1997) pp.1072--1078.
\item {44.}
D. M. McAvity and H. Osborn, {\it Asymptotic expansion of the heat 
kernel for generalized boundary conditions}, Class. Quantum Grav.
{\bf 8} (1991) pp.1445--1454.
\item {45.}
J. S. Dowker and K. Kirsten, {\it Heat-kernel coefficients for 
oblique boundary conditions}, Class. Quantum Grav. {\bf 14} (1997)
pp.L169--L175.
\item {46.}
J. S. Dowker and K. Kirsten, {\it The $a_{3/2}$ heat-kernel coefficient 
for oblique boundary conditions}, Class. Quantum Grav. {\bf 16}
(1999) pp.1917--1936.
\item {47.}
A. O. Barvinsky, {\it The wave function and the effective action in
quantum cosmology: covariant loop expansion}, Phys. Lett. B {\bf 195}
(1987) pp.344--348.
\item {48.}
G. Esposito, A. Yu. Kamenshchik, I. V. Mishakov and G. Pollifrone,
{\it One-loop amplitudes in Euclidean quantum gravity}, Phys. Rev.
D {\bf 52} (1995) pp.3457--3465.
\item {49.}
H. C. Luckock, {\it Mixed boundary conditions in quantum field
theory}, J. Math. Phys. {\bf 32} (1991) pp.1755--1766.
\item {50.}
P. Di Vecchia, A. Liccardo, R. Marotta and F. Pezzella,
{\it Gauge/gravity correspondence from open/closed string duality},
JHEP 0306 (2003) 007.

\bye